\documentstyle[12pt,psfig]{article}
\begin{document}
\baselineskip 18 true pt
\parskip=4 true pt
\newcommand{\be}{\begin{equation}}
\newcommand{\ee}{\end{equation}}
\newcommand{\beq}{\begin{eqnarray}}
\newcommand{\eeq}{\end{eqnarray}}
\newcommand{\bear}{\begin{array}}
\newcommand{\ear}{\end{array}}
\newcommand{\D}{\displaystyle}
\thispagestyle{plain}
\title{DIRAC CURRENT IN EXPANDING SPACETIME}
\author{N.G.Sarkar $^{a)}$, S.Biswas $^{*a),b)}$ and A.Shaw $^{**a)}$ \\
a) Department of Physics, University of Kalyani, West Bengal,\\ 
India, Pin.- 741235 \\
b) IUCAA, Post bag 4, Ganeshkhind, Pune 411 007, India\\
 $*$ e-mail : sbiswas@klyuniv.ernet.in\\
 $**$ e-mail : amita@klyuinv.ernet.in} 
\date{}
\maketitle
\begin{center}
\parbox{12cm}{ABSTRACT: We study the behaviour of Dirac current in expanding spacetime
with Schr{\"o}dinger and de Sitter form for the evolution of the
scale-factor. The study is made to understand the particle-antiparticle
rotation and the evolution of quantum vacuum leading to particle
production in such spacetime.\\[0.5cm]
}
\end{center}

\baselineskip=18pt
\section{Introduction}
The study of particle production is still an active research, especially
in early universe inflationary cosmology. The non-linear dynamics of
quantum fields in fixed cosmological backgrounds reveal new phenomena
when compared to a Minkowski analysis. Such an analysis allows to obtain
the metric dynamically from the quantum fields propagating in that
metric. The dynamics of the scale factor $a(t)$ is driven by semiclassical
Einstein equations
\be
\frac{1}{8\pi G_{_{R}}}G_{_{\mu\nu}}+\frac{\Lambda_{_{R}}}{8\pi G_{_{R}}}
\,g_{_{\mu\nu}}+({\rm higher\,\,
curvature})=-\left<T_{_{\mu\nu}}\right>_{_{R}}.
\ee
Here $G_{_{R}}$, $\Lambda_{_{R}}$ are the normalized values of Newton's
constant and the cosmological constant respectively and $G_{_{\mu\nu}}$
is the Einstein tensor. The higher curvature terms are included to absorb
ultraviolet divergences. Recently [1] we solved equation (1) numerically
taking in the right hand side of Eq.(1) the contributions due to one
loop
quantum corrections plus the energy density due to particle production.
The energy density of the produced particles was calculated using the
method of the complex time WKB approximation (CWKB) and the energy
density so calculated through particle-antiparticle rotation turns out
to be $\rho_{_{r}}/a^4$, as if  it acts like a radiation density term.
The result is quite surprising. However, the numerical solution of (1)
reveals that the universe evolves in a de Sitter phase, avoids
singularity and at late time behaves like a radiation dominated or
matter dominated universe depending upon the mode of particle
production. It is now believed that the parametric and spinodal
instabilities give rise massive particle production and is a result
of coherent oscillations of inflaton as it evolves in time. The
equilibrium dynamics or out of equilibrium dynamics, whichever one
we prefer to use in (1), we need solutions of Dirac equation
(considering only spin-$\frac{1}{2}$ fields) and solve numerically equation
(1). Such kind of numerical solutions considering scalar field has been carried out by Anderson
[2-5], evaluating $\left<T_{_{\mu\nu}}\right>$ for massless and massive
fields with a chosen vacuum for quantum fields. For out of equilibrium
situation, one evaluates $\left<T_{_{\mu\nu}}\right>$ for any state
$\left\vert\right>$ such that the quantum field operator $\Phi(x)$ is
written as
\be
\Phi(x)=\phi(x)+\psi(x),
\ee
where $\psi(x)$ is a new quantum operator. Here
\be
\phi(x)=\left<\Phi(x)\right>,
\ee
\be
\left<\psi(x)\right>=0.
\ee
For a given Lagrangian using Hartree or large $N$ approximation one
evaluates $\phi(x)$ and $\psi(x)$ and calculates $\epsilon$ and $P$
using renormalized equations of motion for dynamical evolution (for
details please see [6-12]. We intend to apply this
technique considering $\Phi(x)$ as spinor field. Leaving aside the
fluctuation term $\psi(x)$ in (2), we study in this work the evolution
of the fields on a fixed FRW background. Before going into the text of the 
paper we note that the numerical solutions of (1), with massless conformaly 
coupled scalar fields contributing to $< T_{\mu\nu} >$ through one loop quantum 
corrections [see ref. 1-5] supposed to be valid also in spinor case when 
curvature is much greater than the mass of the field, reveal that the universe 
evolves in a deSitter phase, bounces and then emerges into classical universe, 
depending upon the matter content to be fixed by the mechanism of particle 
production in the model. In the literature, though most works concentrate on 
scalar fields, the effect of spinor particle production contributing to 
$< T_{\mu\nu} >$ are not well tackled. In our view at a very early stage 
consideration of spinor particle production would be important if one has to 
investigate the preheating and reheating mechanism of early universe, where 
reheating (a process that starts when inflation ends) temperature is solely 
determined by the mechanism of particle production. The numerical investigation 
is no doubt a way to understand and study this aspect but it is a hard nut to 
crack. The analytical treatment allows one to channel the numerical work and 
this is what we intend to take for our discussion. This is also a reason to 
keep $\Lambda_R$ term in equation (1). For further details the reader is 
referred to [2-5], and for numerical solutions backed by analytical input see 
ref. [1]. 
\par
Our aim is now to study the
evolution with scale factor $a(t)={(\frac{t}{t_{_{0}}})}^n$. The cases
$n=\frac{1}{2}$, $n=\frac{2}{3}$ and $n=1$ correspond respectively to
radiation, matter dominated and Schr{\"o}dinger model backgrounds. The
calculations have been carried out in all these backgrounds. We report
here the calculations of $n=1$ case and cite also the results of \hbox{de
Sitter} background for comparison. This work is also motivated to find
an answer to the emergence of $\rho_{_{r}}/a^4$ term as mentioned
earlier. We do not intend here to the solution of dynamical back
reaction equation (1), rather concentrate on the role of fields on
particle production scenario according to CWKB [13-18].
\par
Let us now understand the basis of the CWKB as follows. We write down the 
temporal equation for Dirac particle in FRW spacetime in Klein-Gordon form. 
After separating out the spacetime part we determine the turning points as
usual from the resulting one-dimensional Schr{\"o}dinger-type equation in
time variables. The turning points are found to be complex in the conformal
time. The turning points are identified as points where particle turns
back, not in space but in time, which according to Feynmann-Stuckleberg
prescription must be identified as an event of pair production. Now there is
a mathematical analogy between the Dirac equation in a time varying electric
field and the scattering off a potential which varies in space but not in time.
The unitarity relation obtained from charge conservation relates the reflection
and transmission coefficient $R^{\rm c}$ and $T^{\rm c}$ which refer to the
reflection and transmission coefficient not in space but in time. Pair
production is considered as the reflection of a positron from the vacuum and
$R^{\rm c}$ is the pair production amplitude. It is also found that in the CWKB,
the particle production is basically a resonance particle production and the
essence of the phenomenon of resonance is the rotation of a free-particle
solution to a free-antiparticle solution as if the Dirac current shows a
smooth interpolation from $-J$ to $+J$. The problem in the interpolation is
how to choose the initial conditions i.e., the particle and antiparticle
states. To have a clear understanding on how to choose the initial conditions
of quantum fields, we turn to the explicit solution of Dirac equation in
fixed background and construct the current $J^\mu=\bar{\psi}\gamma^\mu\psi$
to look into the turnings of the current (if it exists) with the evolution
of time. This is the main content of the present paper. We find marked
differences for the cases $n=1, \frac{3}{2}, \frac{1}{2}$ when compared
to the de Sitter evolution $a(t)=\exp(Ht)$.
\par
Though the method of CWKB was originally proposed by us in the context of the 
behaviour of quantum field in curved spacetime, recently many workers [19-21]
are studying this aspect. Recently [22] we have established the rotation of 
currents that we are advocating during particle production and find that the 
CWKB results are gauge invariant result akin to Schwinger's gauge invariant 
result related to particle production. The present work is an attempt to study 
this `rotation of current' in the context of curved spacetime, in order to 
understand the interplay of Minkowski and Rindler vacuum in relation to 
particle production. 
\par
The plan of the paper is organized as follows. In section {\bf 2} we discuss
the Dirac equation in expanding spacetime. In section {\bf 3} we describes 
the basics of CWKB. In section {\bf 4} we discuss the
solutions for (i) $a(t)=a_{_{0}}t$ and (ii) $a(t)=\exp(Ht)$ along with the
corresponding currents. In section {\bf 5}, we discuss the numerical results.
The de Sitter case is discussed with a view to understand the emergence of
radiation dominated universe from the act of particle production in section
{\bf 6}. We end up with a concluding section.    
\section{The Dirac Equation in Expanding Non-Flat Spacetime}
In curved spacetime the Dirac equation is taken to be  
\be
\left[i\gamma^{\mu}(x)\partial_{\mu}-i\gamma^{\mu}(x)
\Gamma_{\mu}(x)\right]{\Psi}=m{\Psi}.
\ee
where $\gamma^\mu(x)$ are the curvature dependent Dirac matrices,
$\Gamma_\mu
(x)$ are the spin connections and $m$ is the mass of the particle.
\par
The metric, we shall consider here, the Robertson-Walker type
\be
ds^2=dt^2-a^2(t)\left[{\frac{dr^2}{1-kr^2}}+r^2
d\theta^2+r^2\sin^2\theta\,d\varphi^2\right],
\ee
that describes a homogeneous isotropic non-flat universe, where
$k=0, +1, -1$, corresponds to Euclidean space, a spherical space and a
pseudo-spherical space respectively. From Eq.(6) we may write the 
metric tensor as
\be
g_{\mu\nu}=\left( \bear{cccc}
1 & 0 & 0 & 0 \\[1mm] 0 & -{\D{\frac{a^2}{\rho^{\,2}}}}
& 0 & 0 \\[2mm] 0 & 0 & -{a^2 r^2} & 0 \\ 0 & 0 & 0 & -{a^2 r^2 \,
\sin^2\theta}\ear \right) 
\ee
and its inverse will be of the form
\be
g^{\mu\nu}=\left(\bear{cccc}
1 & 0 & 0 & 0 \\[2mm] 0 & -{\D\frac{\rho^2}{a^2}}
& 0 & 0 \\[2mm] 0 & 0 & -{\D\frac{1}{a^2 r^2}} & 0 \\[2mm]
 0 & 0 & 0 & -{\D\frac{1}{
a^2 r^2 \,\sin^2\theta}} \ear \right),
\ee
with $\rho^2 =1-kr^2$.
\par
The affine connections $\Gamma^{\alpha}_{\mu\nu}$ with $\alpha\,,\,
\mu\,,\,\nu\,=\,0,\,1,\,2,\,3$ calculated using Eqs.(7) and (8) are
as follows:
\be
\Gamma^0_{\mu\nu}=\left( \bear{cccc}
0 & 0 & 0 & 0 \\[2mm] 
0 & {\D\frac{a\dot{a}}{\rho^2}} & 0 & 0 \\[2mm]  
0 & 0 & {a\dot{a}r^2} & 0 \\[2mm]  0 & 0 & 0 & 
{a\dot{a}r^2 \,\sin^2\theta} \ear \right),
\ee
\be
\Gamma^1_{\mu\nu}=\left(\bear{cccc}
0 & {\D\frac{\dot{a}}{a}} & 0 & 0 \\[2mm] 
{\D\frac{\dot{a}}{a}} & {\D\frac{k\,r}{\rho^2}} & 0 & 0 \\[2mm]
0 & 0 & -{\rho^2 r} & 0 \\[2mm]  0 & 0 & 0 & -{\rho^2 r
\,\sin^2{\theta}} \ear \right),
\ee
\be
\Gamma^2_{\mu\nu}=\left( \bear{cccc}
0 & 0 & {\D\frac{\dot{a}}{a}} & 0 \\[2mm] 
0 & 0 & {\D\frac{1}{r}} & 0 \\[2mm]  {\D\frac{\dot{a}}{a}} & 
{\D\frac{1}{r}} & 0 & 0 \\[2mm] 
0 & 0 & 0 & -{\sin\theta \,\cos\theta} \ear \right),
\ee
\be
\Gamma^3_{\mu\nu}=\left( \bear{cccc}
0 & 0 & 0 & {\D\frac{\dot{a}}{a}} \\[2mm] 
0 & 0 & 0 & {\D\frac{1}{r}} \\[2mm]  0 & 0 & 0 & \cot\theta \\[2mm]
{\D\frac{\dot{a}}{a}} & {\D\frac{1}{r}} & \cot\theta & 0  
\ear\right).
\ee
Eq.(6) may also be written as
\be
ds^{2}=L^{\alpha}_{\mu} L_{\alpha\nu}\,dx^{\mu}\,dx^{\nu}, 
\ee
where the vierbein-components are
\be
L^{\alpha}_{\mu}=\bear{l}
\mu\rightarrow \\[-1.5mm] \alpha \\ \downarrow \\[2.5cm] \ear
\hspace{-7mm}\left( \bear{cccc}
1 & 0 & 0 & 0 \\[2mm]  
0 & {\D\frac{a}{\rho}} & 0 & 0 \\[2mm]  0 & 0 & {ar} & 0 \\[2mm] 
0 & 0 & 0 & {ar\,\sin\theta} \ear \right).
\ee
We may find out the relations between curvature-dependent Dirac matrices
$\gamma_{\mu} (x)$ and curvature-independent Dirac matrices 
$\gamma_{_{\alpha}}$, having relations $\gamma_{_{\mu}}\gamma_{_{\nu}}+
\gamma_{_{\nu}}\gamma_{_{\mu}}=2g_{_{\mu\nu}}$ and $\gamma_{_{a}}\gamma_{_{b}}+
\gamma_{_{b}}\gamma_{_{a}}=2\eta_{_{ab}}$ respectively with $\mu,\,\nu=0, 1, 2,
3$ and $\eta_{_{00}}=1,
\,\eta_{_{ii}}=-1,\,\eta_{_{ij}} =0 $ for $i\ne j$, with the
help of the vierbeins $L^{\alpha}_{\mu}$ by
\be
\gamma_{\mu}(x)=L^{a}_{\mu}\gamma_{a},
\ee
and henceforth the curvature-dependent Dirac matrices $\gamma^{\mu}
(x)$ may be calculated from the relations 
\be
\gamma^{\mu}(x)=g^{\mu\nu}\gamma_{\nu}(x).
\ee
\par
Two sets of curvature-dependent Dirac matrices through the Eqs.
(15) and (16) obtained in terms of curvature-independent Dirac matrices
are as follows:
\be
\gamma_{_{0}}(x)=\gamma_{_{0}},\;\;
\gamma_{_{\,1}}(x)={\D\frac{a}{\rho}}\gamma_{_{1}},\;\; 
\gamma_{_{2}}(x)=ar\gamma_{_{2}},\;\;
\gamma_{_{3}}(x)=ar\,\sin\theta \,\gamma_{_{3}}, 
\ee
and
\be
\gamma^{0}(x)=\gamma_{_{0}},\;\, 
\gamma^{1}(x)=-{\D\frac{\rho}{a}}\gamma_{_{1}},\;\,
\gamma^{2}(x)=-{\D\frac{1}{ar}}\gamma_{_{2}},\;\,
\gamma^{3}(x)=-{\D\frac{1}{ar\,\sin\theta}} \,\gamma_{_{3}}.  
\ee
The spin connections are given by the equation
\be
\Gamma_{\mu}=-{\frac{1}{8}} \left[\gamma^{\nu}(x),
{\gamma_{\nu} (x)}_{;\,\mu} \right], 
\ee
where, the covariant derivative is denoted by ``semicolon (;)".
Also using Eqs.(9), (10), (11), (12), (17) and (18) we find the spin connections
through Eq.(19) as
\be
\bear{l}
\Gamma_{_0}=0, 
\\[2mm] \Gamma_{_1}={\D\frac{1}{2}}{ \,\D\frac{\dot{a}}{\rho}}\gamma_{_0}
\gamma_{_1}, 
\\[2mm] \Gamma_{_2}={\D\frac{1}{2}} \left(\dot{a}r\gamma_{_0}
\gamma_{_2}-\rho\gamma_{_1}\gamma_{_2}\right), 
\\[2mm] \Gamma_{_3}={\D\frac{1}{2}} \left(\dot{a}r\gamma_{_0}
\gamma_{_3}\sin\theta-\rho\gamma_{_1}\gamma_{_3}\sin\theta
-\gamma_{_2}\gamma_{_3} \cos\theta\right). 
\ear
\ee
Thus the Dirac equation (5) takes the form through the Eqs.(18) and (20)
as
\beq
\Big[i\gamma_{_0}{\partial_t}+i{\frac{3}{2}}{\frac{\dot{a}}{a}}
\gamma_{_0}&-&{\frac{i}{a}}\Big\{\rho\gamma_{_1}\partial_r+
{\frac{1}{r}}\gamma_{_2} \partial_{\theta}+{\frac{1}{r\,\sin\theta}}\,
\gamma_{_3} \partial_{\varphi}\nonumber\\ [2mm]
&+&{\frac{1}{r}}\Big(\rho\gamma_{_1}+{\frac{1}{2}}\gamma_{_2}
\cot\theta\Big)\Big\}-m\Big]{\Psi}=0.
\eeq
Using the unitary transformation $S$ such that 
\be
\Psi=S\psi,
\ee
where
\be 
S={\frac{1}{2}}\left(\gamma_{_1}\gamma_{_2}+\gamma_{_2}
\gamma_{_3}
+\gamma_{_3}\gamma_{_1}+1\right)\,e^{-\frac{1}{2}\,\theta \,
\gamma_{_3}
\,\gamma_{_1}} \,e^{-\frac{1}{2}\,\varphi \,\gamma_{_1} \,\gamma_{_2}},
\ee
and its inverse
\be
S^{-1}= e^{\frac{1}{2}\,\varphi \,\gamma_{_1} \,\gamma_{_2}} \,
e^{{\frac{1}{2}}\,
\theta \,\gamma_{_3} \,\gamma_{_1}} \,{\frac{1}{2}}\left(1-\gamma_{_1}
\gamma_{_2}
-\gamma_{_2}\gamma_{_3}-\gamma_{_3}\gamma_{_1}\right),
\ee
the Dirac equation finally will be of the form
\be
\left[\partial_{t}+\frac{3}{2}\frac{\dot{a}}{a}+im\gamma_{_0}
-{\frac{1}{a}}\left\{\overrightarrow{\alpha}.\overrightarrow{\nabla}
+(\rho-1)\alpha_r\left(\partial_r+{\frac{1}{r}}\right)
\right\}\right]\psi=0,
\ee
where, we have used $\gamma_{_i}=\gamma_{_0}\alpha_i$, $\alpha_i=
\gamma_{_0}\gamma_{_i}$ and the transformation of the curvature-independent
Dirac matrices in spherical polar coordinates by
\be
\left( \bear{c}
\gamma_{_{r}} \\ \gamma_{_{\theta}} \\ \gamma_{{_{\,\varphi}}}
\ear \right)=\left( \bear{ccc}
\sin\theta \,\cos\varphi & \sin\theta \,\sin\varphi & \cos\theta \\
\cos\theta \,\cos\varphi & \cos\theta \,\sin\varphi & -\,\sin\theta \\
-\,\sin\varphi & \cos\varphi & 0 \ear \right)\left( \bear{c}
\gamma_{_{1}} \\ \gamma_{_{2}} \\
\gamma_{_{3}} \ear \right).
\ee
\section{Basic Principles of CWKB}
Let us consider a one dimensional Schroedinger equation, not in space but in 
time,
\be
{{d^2}\over {dt^2}}\psi+\omega^2 (t) \psi=0.
\ee
In CWKB we consider $t$ to be a complex variable and assume $\omega (t)$ has 
complex turning points given by
\be
\omega^2 (t_{1,2}) =0.
\ee
Defining
\be
S(t_f,t_i) = \int^{t_f}_{t_i} \omega (t) dt,
\ee
the solution of (27) in CWKB is written as
\be
\psi (t) 
\matrix{\longrightarrow \cr t\rightarrow \infty} 
\exp{\left[iS(t,t_0)\right]}+R\exp{\left[ -iS(t,t_0)\right]}.
\ee
Here $t_0$ and $t$ are real where $t_0$ is arbitrary  and $t_0>t$. In (30) 
$R$ is given by
\be
R={{-i\exp{\left[2iS(t_1,t_0)\right]}}\over {1+\exp{\left[2iS(t_1,t_2)\right]}}},
\ee
where $t_{1,2}$ are the complex turning points determined from (28). The 
interpretation of (30) and (31) is as follows. In (30) the first term is the 
direct ray. It starts from $t_0>t_1$, moving leftward arrives at a real $t<t_0$. 
The second term in (30) corresponds to reflected part. A wave starting from $t_0$ 
reaches the complex turning point $t_1$ and after bouncing back from $t_1$ 
reaches $t$. It is represented as 
\be
(-i)\exp{\left[ iS(t_1,t_0) -iS(t,t_1)\right]}.
\ee
The contribution (32) is then multiplied by the repeated reflections between 
$t_1$ and $t_2$ and the multiple reflection is written as 
\be
\sum_{\mu =0}^{\infty}\left[ -i\exp{\{iS(t_1,t_2)\}}\right]^{2\mu}=
{1\over {1+\exp{\left[2iS(t_1,t_2)\right]}}}.
\ee
The combined contributions (32) and (33) comprise the second term of (30). 
For convenience we have neglected the WKB pre-exponential factor throughout.
\par
Let us now obtain a dynamical picture of particle production from the foregoing
discussion of reflection in time. Consider a potential $V\sim \exp{(-i\omega t)}$
to supply energy for the pair $(e^+e^-)$ creation. In Feynmann's space-time 
diagram, we represent it as in Fig. 1(a).
\be
V\rightarrow 
(e^+,E_{e^+})\uparrow+(e^-,E_{e^-})\uparrow .
\ee
Here the positron of energy $E_{e^+}$ and the electron of energy $E_{e^-}$ both 
are moving forward in time out of the potential site and is represented by the 
arrow beside them. According to Feynmann-Stuckleberg (F-S) prescription, the 
negative energy particle solution propagating backward in time as being 
equivalent to positive energy anti-particle solution moves forward in time. 
Using this prescription, (see fig. 1(b)) 
\[ (e^-,E_{e^-})\uparrow 
\matrix{F. S \cr \Longrightarrow \cr prescription}
(e^+,-E_{e^-})\downarrow\]
i.e., a negative energy positron $(E_{e^+} = -E_{e^-})$ moving backward in time 
(represented by down arrow) meets the potential site $V$ (that acts as turning 
points) and moves forward in time. Thus F-S prescription applying to (34) now reads 
\[ (e^+,E_{e^+})\uparrow+(e^-,E_{e^-})\uparrow 
\matrix{F. S \cr \Longrightarrow \cr prescription}
(e^+,E_{e^+})\uparrow+(e^+,-E_{e^-})\downarrow .\]
Thus pair production can be viewed as a process of reflection in time. For further 
details the reader is referred to [18] and also [13] and [14] to find the 
equivalence,
\[ \vert {\rm{pair\;\; production\;\; amplitude}}\vert = 
\vert {\rm{Reflection\;\; amplitude}}\vert .\] 
\par
If we now consider $\exp{\left[+iS(t,t_0)\right]}$ as antiparticle solution, the 
reflected component is interpreted as a particle moving forward in time. This is 
the Klein paradox-like situation not in space but in time and $R$ is interpreted 
as pair-production amplitude. The essence of (30) is that there is no particle 
at $t\rightarrow -\infty$,i.e.,
\be
\psi_{in}               
\matrix{\longrightarrow \cr t\rightarrow -\infty} 
\exp{\left[ iS(t,t_0)\right]}, \qquad t<0
\ee 
but at $t\rightarrow +\infty$, (35) evolves into 
\[ \psi_{in}               
\matrix{\longrightarrow \cr t\rightarrow +\infty} 
\exp{\left[ iS(t,t_0)\right]} +R\exp{\left[-iS(t,t_0)\right]}.\]
We applied the results (30) and (31) in various expanding spacetime [12-14] with 
remarkable results.
\section{Solutions of Dirac Equation}
In case of flat spacetime i.e. $k=0(\rm{or} \,\rho=1)$, we will, here,
find out the solutions of the Dirac Eq.(25) and it will take its form as  
\be
\left[\partial_t+{\frac{3}{2}}{\frac{\dot{a}}{a}}
+im\gamma_{_{0}}
-{\frac{1}{a}}\overrightarrow{\alpha}.\overrightarrow{\nabla}
\right]\psi=0.
\ee
Let
\be
\psi(x,t)={\D\frac{e^{i{\bf k.x}}}{{(2\pi)}^{3/2}}}
{\left(\bear{c} f({\bf k}, t) \\[2mm] g({\bf k}, t) \ear \right),}
\ee
and thus  we get for two-component spinors $f({\bf k}, t)$ and 
$g({\bf k}, t)$ the coupled equations:
\be
\left(\partial_t+{\frac{3}{2}}{\frac{\dot{a}}{a}}+im\right)f
-{\frac{i}{a}}\vec{k}.\vec{\sigma}g=0,
\ee
and
\be
\left(\partial_t +{\frac{3}{2}}{\frac{\dot{a}}{a}}-im\right)g
-{\frac{i}{a}}\vec{k}.\vec{\sigma}f=0.
\ee
We get in turn the uncoupled equation for $f({\bf k},t)$ as
\be
\left(\partial_t^2-{\frac{1}{2}}{\frac{\ddot{a}}{a}}+{\frac{1}{4}}
{\frac{{\dot{a}}^2}{a^2}}+im{\frac{\dot{a}}{a}}+m^2
+{\frac{k^2}{a^2}}\right)h=0, 
\ee
where we have used
\be
f={\frac{1}{a^2}}h.
\ee
\subsection{Case-I: Schr{\"o}dinger Model}
In Schr{\"o}dinger model, we find 
\be
a(t)=a_{_{0}}t.
\ee
Then Eq.(40) becomes
\be
\left[\partial_{_{t}}^{^{2}}+\frac{k^{^{2}}/a_{_{0}}^{^{2}}+\frac{1}{4}}
{t^{^{2}}}+\frac{im}{t}+m^{^{2}}\right]h=0.
\ee
After changing the variable from $t$ to $z^{^{2}}=-4m^{^{2}}t^{^{2}}$ or
$z=\pm{2imt}$, we get, from Eq.(43),
\be
\left[\frac{d^{^{2}}\,\,}{dz^{^{2}}}-\frac{1}{4}+\frac{\pm{\frac{1}{2}}}{
z}+\frac{\frac{1}{4}-(\pm{ik/a_{_{0}}})^{^{2}}}{z^{^{2}}}\right]h=0,
\ee
which is Whittaker differential equation having two independent solutions:
\be
h_{_{\pm}}(z)=W_{_{\pm{\frac{1}{2}}, \frac{ik}{a_{_{0}}}}}(\pm{2imt}).
\ee
Thus the upper components of the wave function in (37) become
\be
f_{_{\pm}}=\frac{1}{a^{^{2}}}h_{_{\pm}}(z)=\frac{1}{a_{_{0}}^{^{2}}t^{^{2}}}
W_{_{\pm{\frac{1}{2}}, \frac{ik}{a_{_{0}}}}}(\pm{2imt})
\ee
and the lower components will be
\be
g_{_{\pm}}=y_{_{\pm}}\,\frac{1}{k^{^{2}}}\left(\bear{cc}k_{_{3}}&k_{_{-}}\\
k_{_{+}}&-k_{_{3}}\ear\right),
\ee
where
\beq
&&y_{_{\pm}}=-\frac{it}{a_{_{0}}}\left(\partial_{_{t}}+\frac{3}{2t}+im\right)
\frac{1}{t^{^{2}}}W_{_{\pm{1/2},\,ik/a_{_{0}}}}(\pm{2imt}) \nonumber\\[2mm]
&&\hspace{5.6mm}=-\frac{it}{a_{_{0}}}\left(\frac{\pm{2im}}{t^{^{2}}}
{\dot{W}}_{_{\pm}}
-\frac{1}{2t^{^{3}}}W_{_{\pm}}+\frac{im}{t^{^{2}}}W_{_{\pm}}\right)
\eeq
and $k_{_{\pm}}=k_{_{1}}\pm{ik_{_{2}}}$.
But Whittaker functions $W_{_{\pm}}$ satisfy the following identities:
\be
{\dot{W}}_{_{\kappa,\mu}}(z)=\left(\frac{1}{2}+\mu-\kappa\right)\left(
\frac{1}{2}-\mu-\kappa\right)\frac{1}{z}W_{_{\kappa-1, \mu}}+\left(
\frac{\kappa}{z}-\frac{1}{2}\right)W_{_{\kappa, \mu}}
\ee
and
\be
{\dot{W}}_{_{\kappa,\mu}}(z)=-\frac{1}{z}W_{_{\kappa+1, \mu}}-\left(
\frac{\kappa}{z}-\frac{1}{2}\right)W_{_{\kappa, \mu}}.
\ee
We use the first identity for $\kappa=+\frac{1}{2}$, $z=2imt$ and
for second identity $\kappa=-\frac{1}{2}$, $z=-2imt$. Thus Eq.(48) gives,
using Eqs.(49) and (50),  
\be
y_{_{+}}=-\frac{ik^{2}}{a_{_{0}}^{3}}\,\frac{1}{t^{2}}\,W_{_{-\frac{1}
{2}, \frac{ik}{a_{_{0}}}}}(2imt)
\ee
and  
\be
y_{_{-}}=\frac{i}{a_{_{0}}t^{2}}\,W_{_{\frac{1}
{2}, \frac{ik}{a_{_{0}}}}}(-2imt).
\ee
Denoting the constant spinor components of $f$ by $\left(\bear{c}1\\0\ear
\right)$ and
$\left(\bear{c}0\\1\ear\right)$, the four independent solutions of the form
(37) will be
\be
\psi_{_{1}}=N_{_{1}}\frac{e^{i\bf{k.x}}}{(2\pi)^{3/2}}\,\frac{1}{a_{_{0}}
^{2}t^{2}}\left(\bear{c}\quad\left( \bear{c}1\\0\ear\right)
W_{_{\frac{1}{2}, 
\,\frac{ik}{a_{_{0}}}}}(2imt)\\[3mm] -\frac{i}{a_{_{0}}}\left( \bear{c}k_{_{3}}
\\k_{_{+}}\ear\right)W_{_{-\frac{1}{2}, \frac{ik}{a_{_{0}}}}}(2imt)\ear\right),
\ee
\be
\psi_{_{2}}=N_{_{2}}\frac{e^{i\bf{k.x}}}{(2\pi)^{3/2}}\,\frac{1}{a_{_{0}}
^{2}t^{2}}\left(\bear{c}\quad\left( \bear{c}0\\1\ear\right)
W_{_{\frac{1}{2}, 
\frac{ik}{a_{_{0}}}}}(2imt)\\[3mm] -\frac{i}{a_{_{0}}}\left( \bear{r}k_{_{-}}
\\-k_{_{3}}\ear\right)W_{_{-\frac{1}{2}, \frac{ik}{a_{_{0}}}}}(2imt)\ear\right),
\ee
\be
\psi_{_{3}}=N_{_{3}}\frac{e^{i\bf{k.x}}}{(2\pi)^{3/2}}\,\frac{1}{a_{_{0}}
^{2}t^{2}}\left(\bear{c}\quad\left( \bear{c}1\\0\ear\right)
W_{_{-\frac{1}{2}, 
\frac{ik}{a_{_{0}}}}}(-2imt)\\[3mm] \frac{ia_{_{0}}}{k^{^{2}}}\left( \bear{c}
k_{_{3}}
\\k_{_{+}}\ear\right)W_{_{\frac{1}{2}, \frac{ik}{a_{_{0}}}}}(-2imt)\ear\right),
\ee
\be
\psi_{_{4}}=N_{_{4}}\frac{e^{i\bf{k.x}}}{(2\pi)^{3/2}}\,\frac{1}{a_{_{0}}
^{2}t^{2}}\left(\bear{c}\quad\left( \bear{c}0\\1\ear\right)
W_{_{-\frac{1}{2}, 
\frac{ik}{a_{_{0}}}}}(-2imt)\\[3mm] \frac{ia_{_{0}}}{k^{^{2}}}\left( \bear{r}
k_{_{-}}
\\-k_{_{3}}\ear\right)W_{_{\frac{1}{2}, \frac{ik}{a_{_{0}}}}}(-2imt)\ear\right).
\ee
It is interesting to see the asymptotic form of these solutions for large
times. Since $W_{_{\kappa, \mu}}(z)\rightarrow z^{\kappa}e^{-z/2}$, for
$-\frac{3\pi}{2}<{\rm arg}\,z<\frac{3\pi}{2}$, we obtain the normalized 
solutions as follow:
\be
\psi_{_{1}}\simeq \frac{e^{i\bf{k.x}}}{(2\pi)^{3/2}}\,\frac{\sqrt{i}}
{{a_{_{0}}}^{3/2}}\frac{1}{t^{3/2}}
\left(\bear{c}\quad\left( \bear{c}1\\0\ear\right)
\\[3mm] -\frac{1}{2ma_{_{0}}}\left( \bear{c}k_{_{3}}
\\k_{_{+}}\ear\right)\,\frac{1}{t}\ear\right)\,e^{-imt},
\ee
\be
\psi_{_{2}}\simeq \frac{e^{i\bf{k.x}}}{(2\pi)^{3/2}}\,\frac{\sqrt{i}}
{a_{_{0}}^{3/2}}\frac{1}{t^{3/2}}
\left(\bear{c}\quad\left( \bear{c}0\\1\ear\right)
\\[3mm] -\frac{1}{2ma_{_{0}}}\left( \bear{r}k_{_{-}}
\\-k_{_{3}}\ear\right)\frac{1}{t}\ear\right)\,e^{-imt},
\ee
\be
\psi_{_{3}}\simeq \frac{e^{i\bf{k.x}}}{(2\pi)^{3/2}}\,\frac{\sqrt{-i}}
{a_{_{0}}^{5/2}}\frac{k}{t^{3/2}}
\left(\bear{c}{-\frac{1}{2im}}\left( \bear{c}1\\0\ear\right)\frac{1}{t}
\\[3mm] \frac{ia_{_{0}}}{k^{2}}\left( \bear{c}
k_{_{3}}
\\k_{_{+}}\ear\right)\ear\right)\,e^{imt},
\ee
\be
\psi_{_{4}}\simeq \frac{e^{i\bf{k.x}}}{(2\pi)^{3/2}}\,\frac{\sqrt{-i}}
{a_{_{0}}^{5/2}}\frac{k}{t^{3/2}}
\left(\bear{c}{-\frac{1}{2im}}\left( \bear{c}0\\1\ear\right)\frac{1}{t}
\\[3mm] \frac{ia_{_{0}}}{k^{2}}\left( \bear{r}
k_{_{-}}
\\-k_{_{3}}\ear\right)\ear\right)\,e^{imt}.
\ee
Where we have determined the normalization constant $N_{_{i}}$ in such a way 
that asymptotically, i.e., in the flat-space limit, we have the usual 
$\delta({\bf{k-k^{\prime}}})$ normalization of the electron's wave function. The
norm of $\psi$ is defined by
\beq
&&\left(\psi^{k}, \psi^{k^{\prime}}\right)=\int_{_{t}}d^{3}x\,a^{{3}}
(t){\bar\psi}^{k}\gamma^{0}\psi^{k^{\prime}}\nonumber\\[3mm]
&&\hspace{1.85cm}=\int_{_{t}}d^{3}x\,a_{_{0}}^{3}t^{3}
\psi_{_{k}}^{\dagger}\psi_{_{ k^{\prime}}}\nonumber\\[3mm]
&&\hspace{1.85cm}\rightarrow\delta({\bf{k-k^{\prime}}})a^3_{_{0}}
t^{3}{\frac{\sqrt{2im}}{
a_{_{0}}^{2}}}{\frac{\sqrt{-2im}}{a_{_{0}}^{2}}}\frac{N^{2}}{t^{3}}.
\eeq
This procedure gives
\be
\bear{c}N_{_{1}}=N_{_{2}}=\sqrt{\frac{a_{_{0}}}{2m}},\\
N_{_{3}}=N_{_{4}}=\frac{k}{\sqrt{2ma_{_{0}}}}.  
\ear
\ee
\subsection{Case-II: de Sitter spacetime}
In case of de Sitter spacetime, the function $a(t)$
is given by
\be
a(t)=e^{Ht}.
\ee
In a similar way, as in \S$\,$4.1, we get four asymptotic normalized solutions as
\be
\psi_{_1}={\frac{e^{i{\bf{k.x}}}}{{(2\pi)}^{3/2}}}
{{\left[{\frac{\frac{\pi\,k}{2H}}{\cos\left({\frac{i\,\pi\,m}{H}}\right)}}
\right]}^{\frac{1}{2}}}{e^{-2Ht}}
{\left(\bear{c}{\left(\bear{c} 1 \\ 0 \ear\right)J_{\nu}(z)} 
\\[3mm] {\frac{i}{k}}\left(\bear{c}
k_{3} \\  k_{+}\ear\right)J_{\nu-1}(z)\ear\right)},
\ee
\be
\psi_{_2}={\frac{e^{i{\bf{k.x}}}}{{(2\pi)}^{3/2}}}
{\left[{\frac{\frac{\pi k}{2H}}{\cos\left({\frac{i\,\pi\,m}{H}}\right)}}
\right]}^{\frac{1}{2}}{e^{-2Ht}}
{\left(\bear{c}{\left(\bear{c} 0 \\ 1 \ear\right)J_{\nu}(z)} 
\\[3mm] {\frac{i}{k}}\left(\bear{c}
k_{-} \\ {-k_{3}}\ear\right)J_{\nu-1}(z)\ear\right)},
\ee
\be
\psi_{_3}={\frac{e^{i{\bf{k.x}}}}{{(2\pi)}^{3/2}}}
{\left[{\frac{\frac{\pi k}{2H}}{\cos\left({\frac{i\,\pi\,m}{H}}\right)}}
\right]}^{\frac{1}{2}}{e^{-2Ht}}
{\left(\bear{c}{\left(\bear{c} 1 \\ 0 \ear\right)J_{-\nu}(z)} 
\\[3mm] {-{\frac{i}{k}}}\left(\bear{c}
k_{3} \\ {k_{+}} \ear\right)J_{-\nu+1}(z)\ear\right)} 
\ee
and
\be
\psi_{_4}={\frac{e^{i{\bf{k.x}}}}{{(2\pi)}^{3/2}}}
{\left[{\frac{\frac{\pi k}{2H}}{\cos\left({\frac{i\,\pi\,m}{H}}\right)}}
\right]}^{\frac{1}{2}}{e^{-2Ht}}
{\left(\bear{c}{\left(\bear{c} 0 \\ 1 \ear\right)J_{-\nu}(z)} 
\\[3mm] {-{\frac{i}{k}}}\left(\bear{c}
k_{-} \\ {-k_3} \ear\right)J_{-\nu+1}(z)\ear\right)}. 
\ee
\par
We now evaluate the Dirac currents for these cases I and II.
\section*{Dirac Current in Case-I:}
Particle production in Schr{\"o}dinger model may easily be
interpreted from
the nature of the curves of Dirac current in a given spacetime. Using
Eq.(5) 
\be
\left[i\gamma^{^{\mu}}(x)\partial_{_{\mu}}-i\gamma^{^{\mu}}(x)\Gamma_{_{\mu}}
(x)\right]\psi=m\psi,
\ee
in an external field $A_{\mu}$ and its conjugate we can express
the Gordon decomposition of current as follows [23]:
\beq
&&j^{\mu}=\bar{\psi}\gamma^{\mu}\psi={\frac{1}{ 
2m}}\Big[i\bar{\psi} \gamma^{\mu}\gamma^{\lambda}\partial_{\lambda}\psi
-i\partial_{\lambda}{\bar{\psi}}\gamma^{\lambda}\gamma^{\mu}\psi \nonumber \\
&&\hspace{3.5cm}-{\bar{\psi}}
\left\{i\left(\gamma^{\mu}\gamma^{\lambda}\Gamma_{\lambda} + \Gamma_{\lambda}
\gamma^
{\lambda}\gamma^{\mu} \right)+ e A_{\lambda}g^{\lambda\mu}\right\} \psi \Big]
\eeq 
For $k=0$ i.e. in the flat spacetime the metric (6) reduces to
\be
ds^2=dt^2-a^2(t)\left(dx^2_1+dx^2_2+dx^2_3\right),
\ee
and also the space-dependent Dirac matrices (18) get the following forms
\be
\gamma^{0}(x)=\gamma_{_{0}},\;\, 
\gamma^{1}(x)=-{\D\frac{1}{a}}\gamma_{_{1}},\;\,
\gamma^{2}(x)=-{\D\frac{1}{a}}\gamma_{_{2}},\,\;
\gamma^{3}(x)=-{\D\frac{1}{a}}\gamma_{_{3}}.  
\ee
Using this equation (71) we have the current densities in components as
\be
j_{_0}={\frac{i}{2ma}} \nabla_{_{j}} \left(\bar{\psi} \gamma_{_{j}}
\gamma_{_{0}} \psi\right) + {\frac{1}{2m}}{\bar{\psi}}\left[i
\stackrel{\leftrightarrow}{\partial_{_{0}}} + e A_{_{0}}\right]\psi
\ee
and
\beq
&&j_{_k}={\frac{i}{2ma}}\,\partial_{_{t}}\left(
\bar{\psi}\gamma_{_{k}} \gamma_{_{0}} \psi \right) + {\frac{i}
{4 m a^2}} \partial_{_{j}} \left( \bar{\psi} \left[ \gamma_{_{j}}
\,,\,\gamma_{_{k}}\right] \psi \right) \nonumber \\ 
&&\hspace{7.8mm}+ {\frac{1}{2ma^2}}{\bar{\psi}}
\left[i\stackrel{\leftrightarrow}
{\partial_{_{k}}}+e\,A_{_{k}}\right]\psi+
{\frac{3i}{2m}}{\frac{\dot{a}}{a^2}} \bar{\psi} \gamma_{_{k}} 
\gamma_{_{0}}\psi.
\eeq
\par
Now using equations (72) and (73), taking $A_{_{\mu}}=0$,  
we get the same results for current densities $j_{_{0}}$, $j_{_{1}}$,
$j_{_{2}}$, and $j_{_{3}}$ for the solutions 
$\psi_{_{1}}$ and $\psi_{_{2}}$ given in Eqs.(57) and (58). These are as
follows:
\be
j_{_{0}}=C_{_{0}}T^{^{3}}+C_{_{0}}^{^{\prime}}T^{^{5}},\;\,
j_{_{1}}=-C_{_{1}}T^{^{5}},\;\,
j_{_{2}}=-C_{_{2}}T^{^{5}},\;\,
j_{_{3}}=-C_{_{3}}T^{^{5}}
\ee
and for the solutions $\psi_{_{3}}$ and $\psi_{_{4}}$ of Eqs.(59) and (60),
the components of the current densities get their forms as 
\be
j_{_{0}}=C_{_{0}}T^{^{3}}+C_{_{0}}^{^{\prime}}T^{^{5}},\;\,
j_{_{1}}=C_{_{1}}T^{^{5}},\;\,
j_{_{2}}=C_{_{2}}T^{^{5}},\;\,
j_{_{3}}=C_{_{3}}T^{^{5}},
\ee
where $C_{_{0}}=\D\frac{1}{8\pi^{3}a_{_{0}}^{3}}$, $\,C_{_{0}}^{\prime}=
\D\frac{k^{2}}{32\pi^{3}m^{2}a_{_{0}}^{5}}$, $\,C_{_{1}}=
\D\frac{k_{_{1}}}{8\pi^{3}ma_{_{0}}^{5}}$, $\,C_{_{2}}=\D\frac{k_{_{2}}}{8
\pi^{3}ma_{_{0}}^{5}}$ and $C_{_{3}}=\D\frac{k_{_{3}}}{8\pi^{3}m
a_{_{0}}^{5}}\,$ with $\,T=\frac{1}{t}$. 
\section*{Dirac Current in Case-II:}
Similarly the four components of the current densities, in de Sitter spacetime,
will get their
forms as
\be
\bear{l}
j_{_{0}}=C_{_{0}}T^{^{3}} \\[2mm]
j_{_{1}}=C\left(k_{_{1}}\sinh{\frac{\pi m}{H}}-k_{_{2}}\sin{T}\right)\,T^{^{4}} 
\\[2mm] 
j_{_{2}}=C\left(k_{_{2}}\sinh{\frac{\pi m}{H}}+k_{_{1}}\sin{T}\right)\,T^{^{4}} 
\\[2mm]
j_{_{3}}=Ck_{_{3}}T^{^{4}}{\sinh{\frac{\pi m}{H}}} 
\ear
\ee
or in brief,
\be
\bear{l}
j_{_{0}}=C_{_{0}}T^{^{3}} \\[2mm]
j_{_{1}}=C_{_{1}}\left(1-D\sin{T}\right)\,T^{^{4}} 
\\[2mm] 
j_{_{2}}=C_{_{2}}\left(1+D_{_{1}}\sin{T}\right)\,T^{^{4}} 
\\[2mm]
j_{_{3}}=C_{_{3}}T^{^{4}} 
\ear
\ee
with the solution $\psi_{_{1}}$ and
where, $T={\frac{2k}{H}}\,e^{-Ht}$, $C_{0}={\frac{H^{3}}{64{\pi}^{3}k^{3}}}$,
$C={\frac{H^4}{128{\pi}^{3}k^{5}\cos\left({\frac{i\,\pi\,m}{H}}\right)}}$,
$C_1=C{k_{1}\sinh({\pi}m/H)}$, $C_{2}=C{k_{2}\sinh({\pi}m/H)}$,
$C_3=C{k_{3}\sinh({\pi}m/H)}$, $D={\frac{k_{2}}{k_{1}\sinh({\pi}m/H)}}$ and
$D_1={\frac{k_1^2}{k_2^2}}D$.  
\par
Let us mention some arguments for citing the Gordon decomposition, though it is 
not required to obtain the expression of currents in the present work.
We also calculated $\bar{\psi}\gamma^{\mu}\psi$ using the exact solutions of 
Dirac equation and find that 
this result obtained by us is coincident with the Gordon decomposition (73). In 
Barut's work [23] this Gordon decomposition is wrongly calculated and has forced us 
to check this result. The Gordon decomposition is carried out to investigate the 
effect of spin towards the rotation of currents which is extremely important in 
understanding the particle production leaving aside the complicacy of vacuum 
state definition in curved spacetime. 
\section{Interpretation of Currents}
In order to interpret the currents qualitatively, we carry out the
numerical analysis. For Schr{\"o}dinger model as well as de Sitter
spacetime we observe that $j_{_{0}}=\bar{\psi}\gamma_{_{0}}\psi$ varies
as $\frac{1}{a^3}$. For the case-I, the nature of the curves  
for $\,j_{_{1}}\,$, $j_{_{2}}\,$, $j_{_{3}}\,$ are of the same 
form for the 
set of solutions $\,\psi_{_{1}}\,$ and $\,\psi_{_{2}}\,$. Similar results are
also obtained for $\psi_{_{3}}$ and $\psi_{_{4}}$. We observe that
$j_{_{i}}(\psi_{_{1, 2}})=-j_{_{i}}(\psi_{_{3, 4}})$, 
though $j_{_{0}}$ remains the
same for both the sets. For convenience
we take $\,C_{_{0}}=C_{_{0}}^{\prime}=C_{_{1}}=C_{_{2}}=C_{_{3}}=1\,$ and 
the curves are reproduced in figure-(2a) for $\,j_{_{0}}\,$, figure-(2b) for 
the first
set of solutions and figure-(2c) for the second set of solutions. The curves in
figure-(2a) and figure-(2c) are of the same nature.
\par
The current in Schr{\"o}dinger model has an interesting properties.
From the figure-(2b) and figure-(2c) it is clear that the current of 
$\psi_{_{1}}$ and $\psi_{_{2}}$ is negative whereas for $\psi_{_{3}}$
and $\psi_{_{4}}$
it is positive. So from equations (57) to (60) it is evident that
the turning (also shown in figure-(2d)) $(\psi_{_{1}},
\psi_{_{2}})\rightarrow(\psi_{_{3}}, \psi_{_{4}})$ would lead to 
particle production. Though the situation here
is less attractive and interpretative compared to de Sitter spacetime,
the plot of current density versus time provides interesting way to 
understand particle production through particle-antiparticle rotation.
Though we do not get a definite conclusions, we believe that the particle
production should somehow be dependent on initial states and the initial 
conditions would greatly effect particle production as well as the evolution
of the universe.
It is worthwhile to mention that the behaviour of current in radiation
dominated universe is very much similar to Schr{\"o}dinger model. It is 
evident from the curves that if there are some fluctuations that take
the system out of equilibrium there will be oscillations between curves
I and II in fig.-2(d). 
\par
In case of de Sitter spacetime the forms  
of $j_{0}$, $j_{_1}$, $j_{_2}$, and $j_{_3}$ are of the same form 
as that of Eq.(77) when we consider any one of the solutions $\psi_{_{\,2}}$, 
$\psi_{_{\,3}}$ or $\psi_{_{\,4}}$ separately. Though there is some differences  
in constants,
only we look into the behaviour of currents with 
time. The nature of the curves are of 
the same form when 
the current density $j_{_{1}}$, $j_{_{2}}$, or $j_{_{3}}$ are plotted
against time $\,t\,$. For convenience we take $C_{0}=C_{1}=C_{2}=C_{3}=D=D_{1}
=1$, $2k=H=1$ and the curve is reproduced in the 
figure-(2e). The oscillating time-dependent current densities 
$j_{_{1}}$ and $j_{_{2}}$ always remain symmetrical about 
the exponentially decaying one $j_{_{3}}$. The 
nature of the curve of $j_{_{0}}$
is of the same form as that of $j_{_{3}}$, not shown in the figure. A single 
oscillation indicates particle 
production (though the nature of the vacuum is unspecified). Repeated 
oscillations indicates particle production \hbox{(at 2a, 2b, 2c, ...)} and 
annihilation (at 2d, 2e, 2f, ...) in course of time. Without going into the details
of quantum vacuum, it can be said that repeated oscillation is an indication
of generation of photon along with particle production. The occurrence of 
oscillations at $\,t\,$ large negative indicate the existence of repeated 
reflected trajectories as is demanded in our {\sc cwkb} approach \hbox{[13,
14,17,18]}. The emergence 
of radiation dominated universe may thus be explained considering production and
annihilation (trough and crest in the Fig.(2e)) as a mode for generation of
photon at late time. In view of our work on de Sitter spacetime [15] the
oscillating behaviour of de Sitter current is very interesting and
worth-pointing in the sense that the emergence of radiation dominated
universe might also also help understand the \hbox{baryon-asymmetry} problem
in early universe.
\par
The numerical parameters adopted in this work are used only to have a quantitative
understanding of photon production in the early universe. We will publish 
shortly the results taking parameters that corresponds to early universe situation.
\section{Photon Dominance in de Sitter Spacetime}
Before we discuss the production of huge number of photons due to particle
production in de Sitter spacetime, let us consider the arguments of Bernido
[24] in this respect. The Green function for a relativistic scalar particle
is expressed as
\be
G(x'',x;m)={(i\hbar)}^{-1}\int_0^\infty\,\exp(-\frac{im^2\lambda}{2})\,K(
x'',x';\lambda),
\ee
where $x=({\bf r},t)$ and $\lambda$ is a time-like variable. Antiparticles in
the path integral approach are identified by particles which move
backwards in ordinary time such that $\partial{t}/\partial{\lambda}<0$,
with $\lambda$ always increasing. Pair production is understood as
follows. Let $A({\bf r},t,\lambda=0)$, $B({\bf r}',t,\lambda=\lambda_{_{
B}})$ and $C({\bf r}_{_{s}},t_{_{s}},\lambda)$ be three spacetime points.
Let a particle leave the point $A$ and move backwards in time, scatter
at $C$, turn back and move forward in time to arrive at $B$. Here
${\bf r}<{\bf r}_{{s}}<{\bf r}'$ and $t_{_{s}}<t$. We note that here the
path from $A$ to $B$ represents a particle moving backward in ordinary
time $t$. In Feynmann's language we interpret it by saying that an
observer at time $t$ sees both a particle and antiparticle coming
from $C$.
\par
In de Sitter spacetime, Eq.(78) has been evaluated in ref.[24]. It
is found that the Green function yields no bound state spectrum
and it exists provided $t''>t'$. It is to be noted that the spacetime
points $x'(={\bf r}',t')$ and $x''(={\bf r}'',t'')$ can be viewed as
the points where a particle and antiparticle can be detected. The
consequence of the constraint $t''>t$ is interpreted as follows.
\begin{description}
\item{(i)} For a given $t$, such that $t'<t<t''$, an observer can only detect
a particle and not its antiparticle. This gives rise to an observed
matter universe.
\item{(ii)} To obtain huge number of photons compared with the baryons
as observed today, consider the particle starting at point $x'=({\bf r}',
t')$. It may follow a path that jogs up and down (with respect to $t$
coordinate) and a huge number of photons is thus generated from pair
annihilation (i.e., creation at the trough and annihilation at the crest).
\end{description}
\par
The arguments placed in (i) and (ii) though seem encouraging but have some
shortfalls. What makes the particle jog up and down? A plausible answer
comes from the CWKB approach where the mechanism described in (i) and
(ii) fit very nicely. The details will be placed elsewhere and here we 
mention the essential aspects. The second order Dirac equation or the
equation (40) shows turning points at
\be
\frac{k}{a}=\pm\,i\left(m+\frac{iH}{2}\right),
\ee
when we put $a(t)=\exp(Ht)$.
\par
According to CWKB, the particle starts from $t_{_{i}}$ and arrives at
$t$ with $t<t_{_{i}}$ without any reflection. We call it direct rays.
There is also a reflected part. A particle starts from $t_{_{i}}$
and moving backwards suffers a reflection at one of the turning points,
say $t_{_{1}}=+i\left(m+\frac{iH}{2}\right)$, and moving forward arrives
at $t$. This is identified as pair production. The reflected part is
enhanced by the repeated reflections between the turning points, in
complex $t$-plane, given by equation (79). As like the arguments of
Bernido [24], the particle jogs up and down resulting in a huge number
of photons. The remarkable equivalence of CWKB with the standard techniques
of particle production has already been established [15,16,18]. In CWKB,
the repeated reflections gives rise to thermal spectrum and a reflection
in time is considered as a rotation of currents i.e., $-J$ to $+J$. In
CWKB, we also find an answer to the Bernido's plausible assumptions. The
matter-antimatter asymmetry can also be explained from our approach.
However, we do not consider it here. Before ending, it is justified
to ask whether the turning points given by (79) have anyway been reflected
in the expression of Dirac currents in the curved spacetime.
\par
Consider the situation just after Big-Bang. We take $k\ll{Ha}$ and $m\ll H$,
the plausible conditions in inflationary cosmology. Under these conditions
the expressions of the currents in Eq.(76) give
\[\frac{k}{a}\simeq\pm\left(\frac{k_{_{2}}}{k_{_{1}}}\right)m,\]
quite consistent with Eq.(79), if we consider the case $m^2<0$ (broken
symmetry). The condition $k\ll Ha$ correspond to long wavelength
fluctuations greater than the Hubble radius, indicating the emergence of
spinodal instabilities, a crude picture of which is reflected in the
expression of Dirac current. However, the situation is not so simple
and straightforward as mentioned. We need here to understand the interplay
of vacuum energy and inflationary universe. The reader is referred to ref.[25]
in this respect. 
\par
It should be pointed out that in a realistic approach, huge photon production 
should follow from high temperature nonperturbative QED, but in curved 
spacetime a complete  theory for which is still lacking. The work of Lotze [26]
may be consulted for simultaneous creation of $e^+e^-$ pairs and photons in RW 
universe. 
\section{Conclusion}
The present work reveals that 
\begin{description}
\item[\qquad(i)] the quantum vacuum at an early epoch starts from a state to be 
characterized at a given time,
\item[\qquad(ii)] during the evolution of quantum vacuum there is rapid particle
production and annihilation, of course in expanding spacetime.
\item[\qquad(iii)] amplitude of oscillation decreases with time indicating a gradual
attainment of stable ground state ( stable vacuum ),
\item[\qquad(iv)] at late time particle production ends.
\end{description}
\par
In view of the above results it would thus be interesting to study the
behaviour of Dirac currents in other expanding spacetimes. Rapid oscillation
in the current is a characteristic feature of inflationary spacetime
and may be introduced through an inflaton field or through a
oscillating
scale-factor (as is found in Starobinsky-type model without phase 
transition). We hope to discuss, in future, particle production, back
reaction, avoidance of singularity [see 1], catastrophic particle-production
through parametric resonance within our framework.

Acknowledgement: The authors are thankful to Dr. Supratic Chakraborty for
his constant encouragement and help during the course of this work. A. Shaw 
acknowledges the financial support from ICSC World Laboratory, LAUSSANE
during the course of the work.
\bigskip
\medskip
\begin{figure}[h]
\psfig{figure=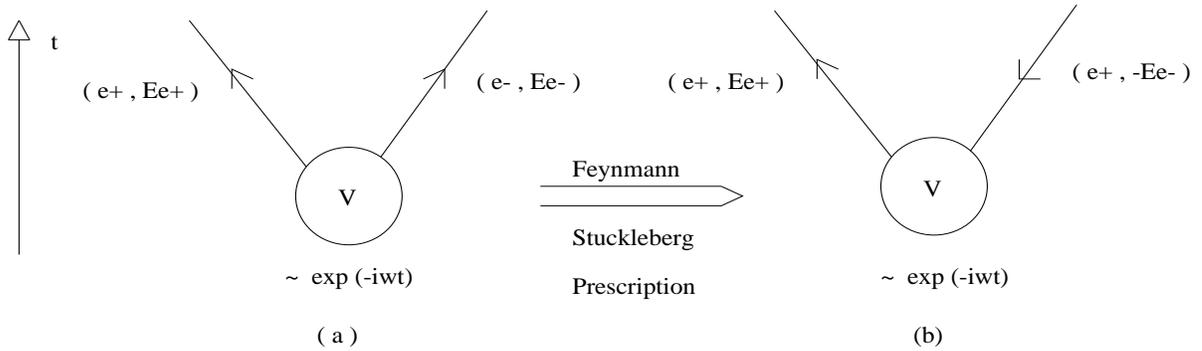,width=16cm,height=4.5cm}
\caption{Feynmann Stuckleberg Prescription} 
\end{figure}
\newpage
\begin{figure}
\vspace{3cm}
\hspace{3.8cm}(a)\\[-7cm]

\psfig{figure=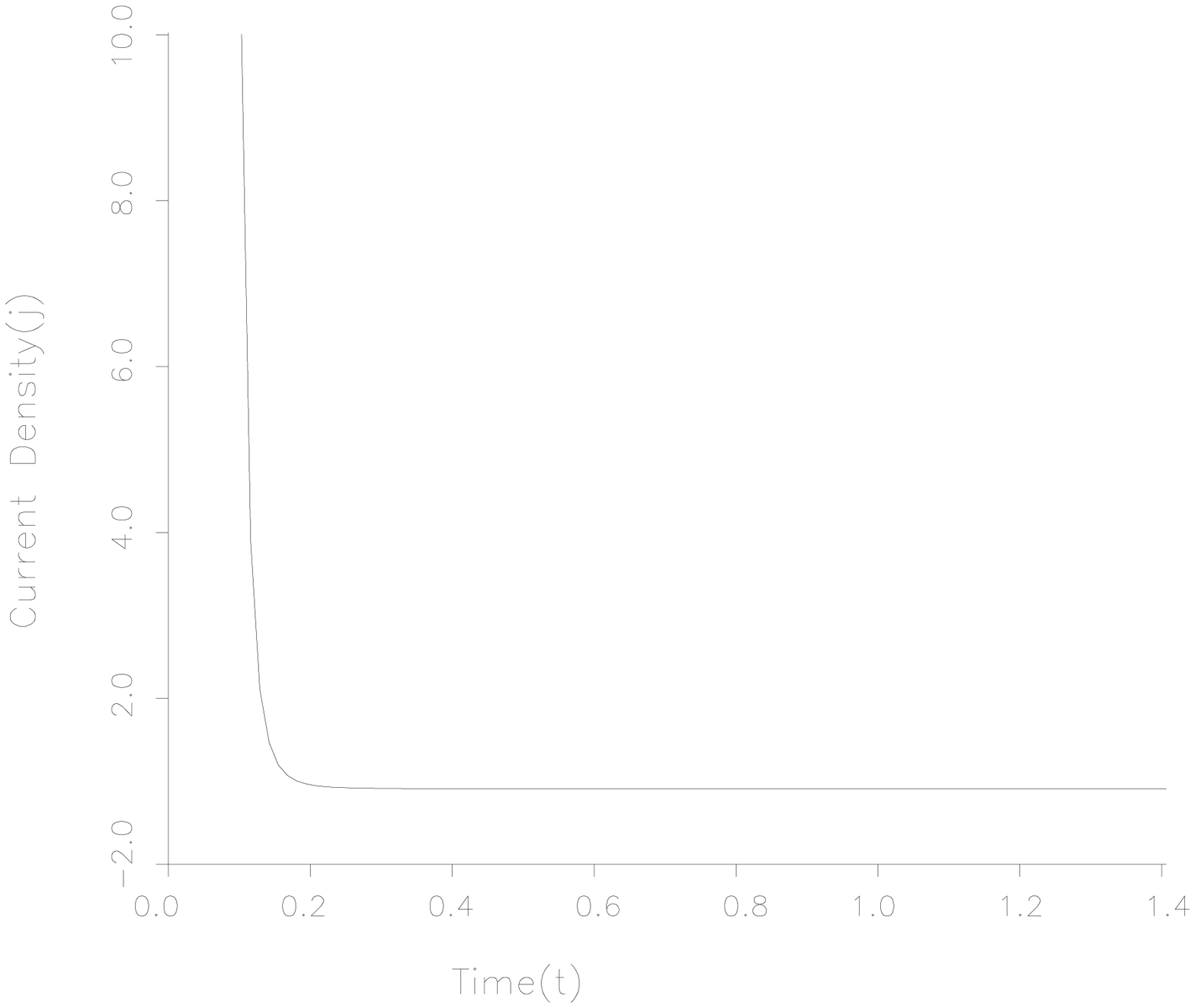,width=7cm,height=6cm}
\end{figure}
\begin{figure}
\vspace{-0.45cm}
\hspace{12.2cm}(b)\\[-7cm]

\hspace{8.4cm}{\psfig{figure=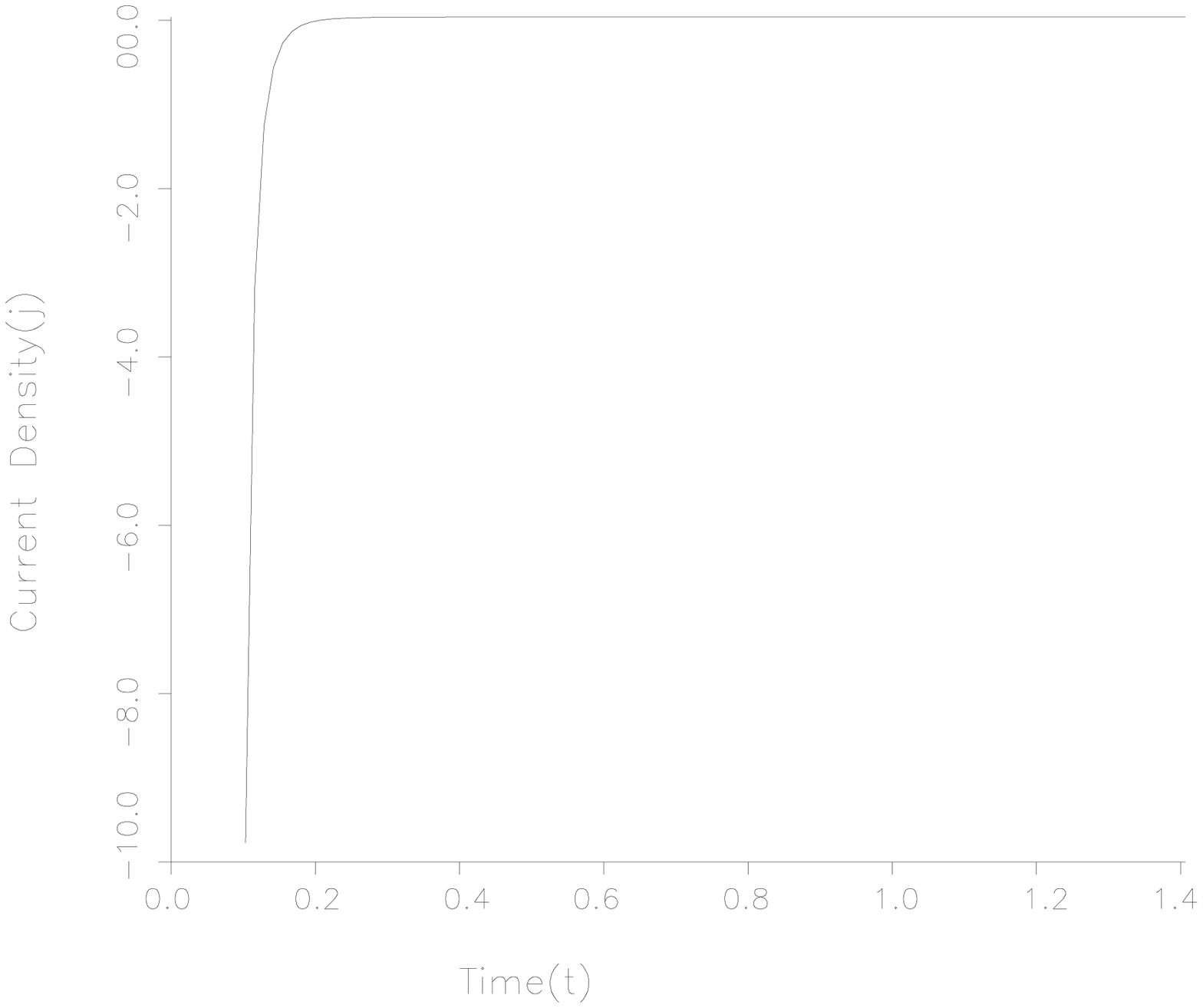,width=7cm,height=6cm}}
\vspace{-4.58cm}
\end{figure}
\begin{figure}
\vspace{12.5cm}
\hspace{3.8cm}(c)\\[-7cm]

\psfig{figure=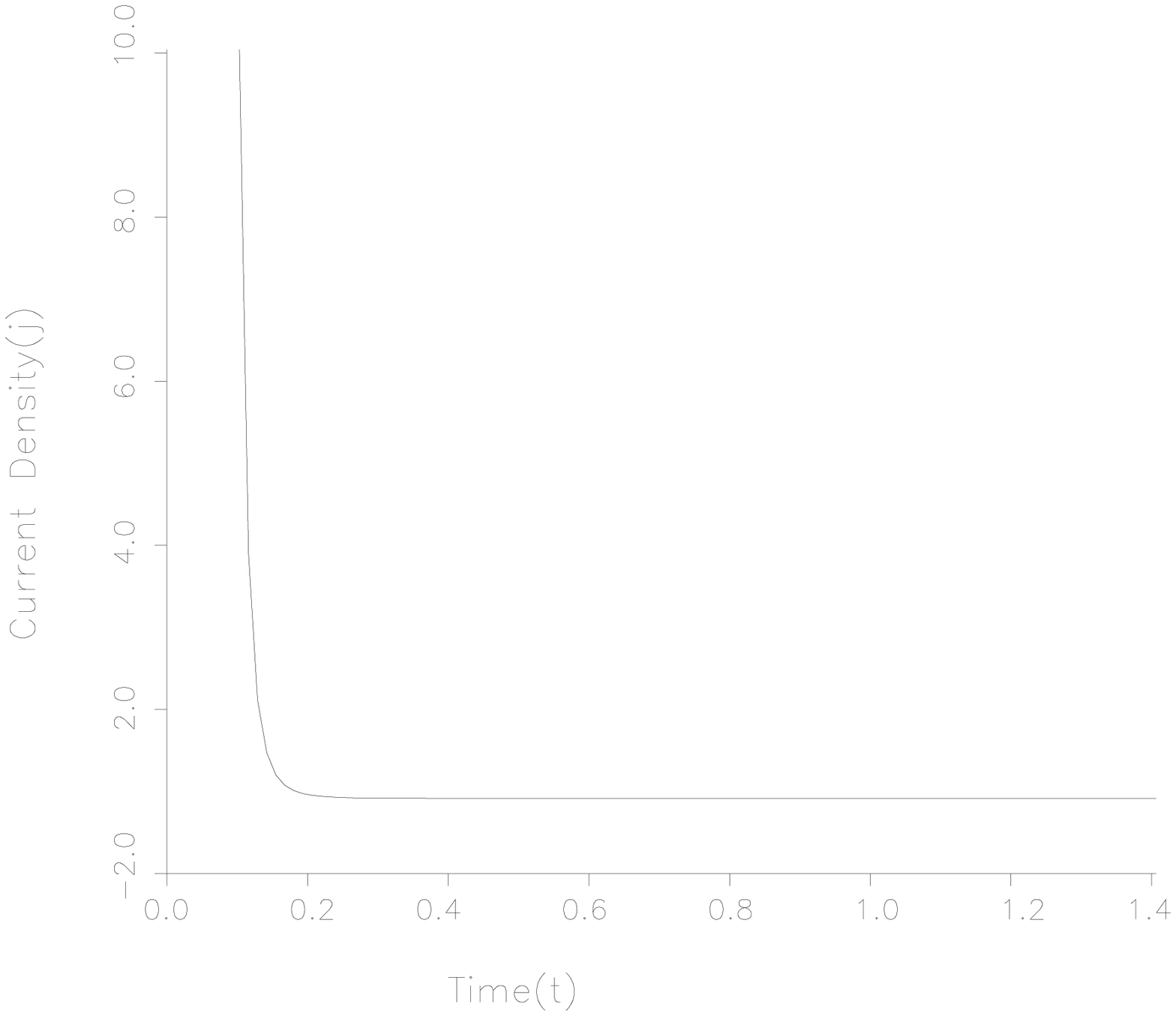,width=7cm,height=6cm}
\vspace{6cm}
\end{figure}
\begin{figure}
\vspace{-6.52cm}
\hspace{12.2cm}(d)\\[-7cm]

\hspace{8.4cm}{\psfig{figure=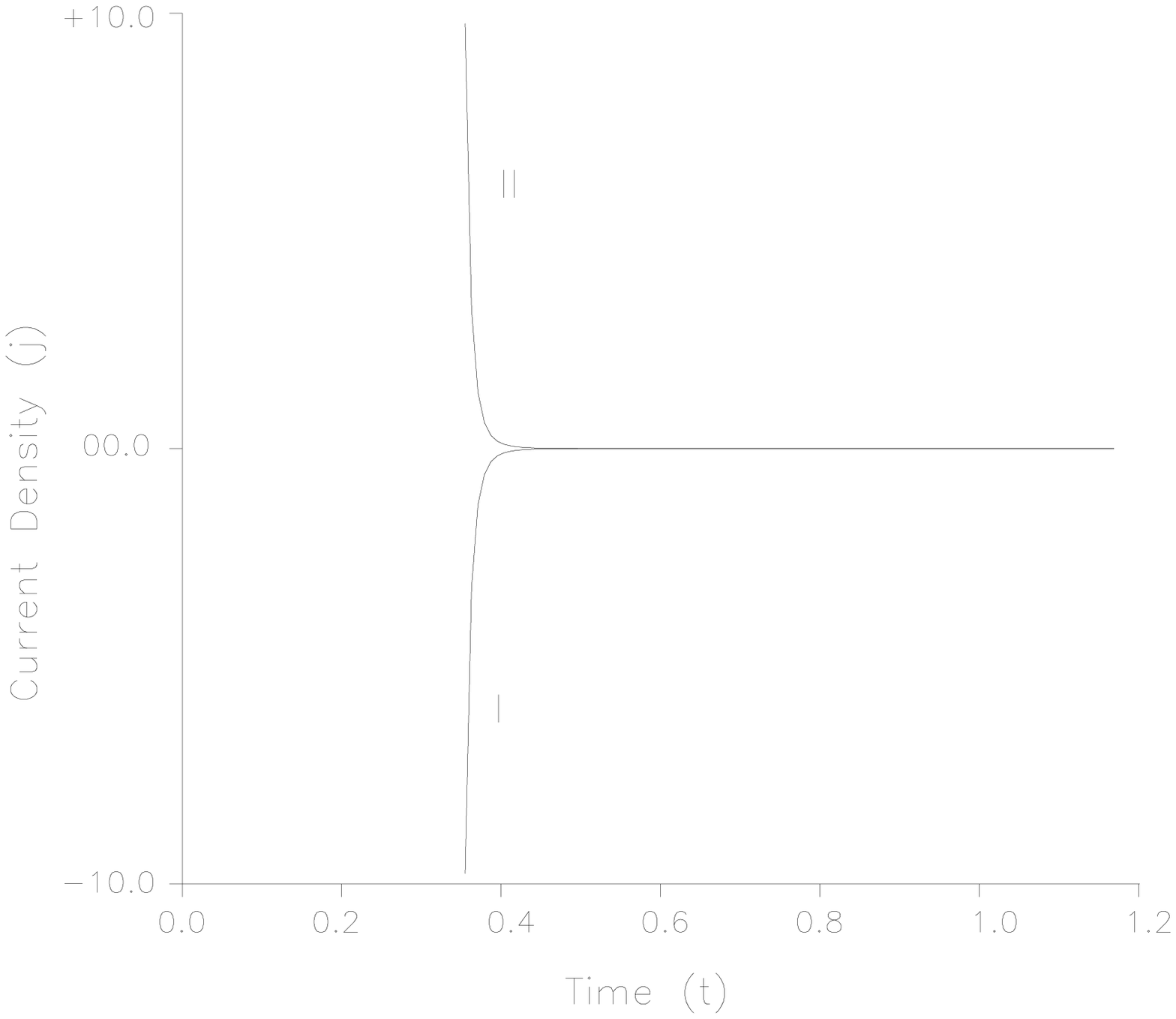,width=7cm,height=6cm}}
\vspace{2.6cm}
\end{figure}
\begin{figure}
\vspace{2cm}
\hspace{2.49cm}{\scriptsize f}\\[1.75cm]
\hspace{3.8cm}(e)\\[-7cm]

\psfig{figure=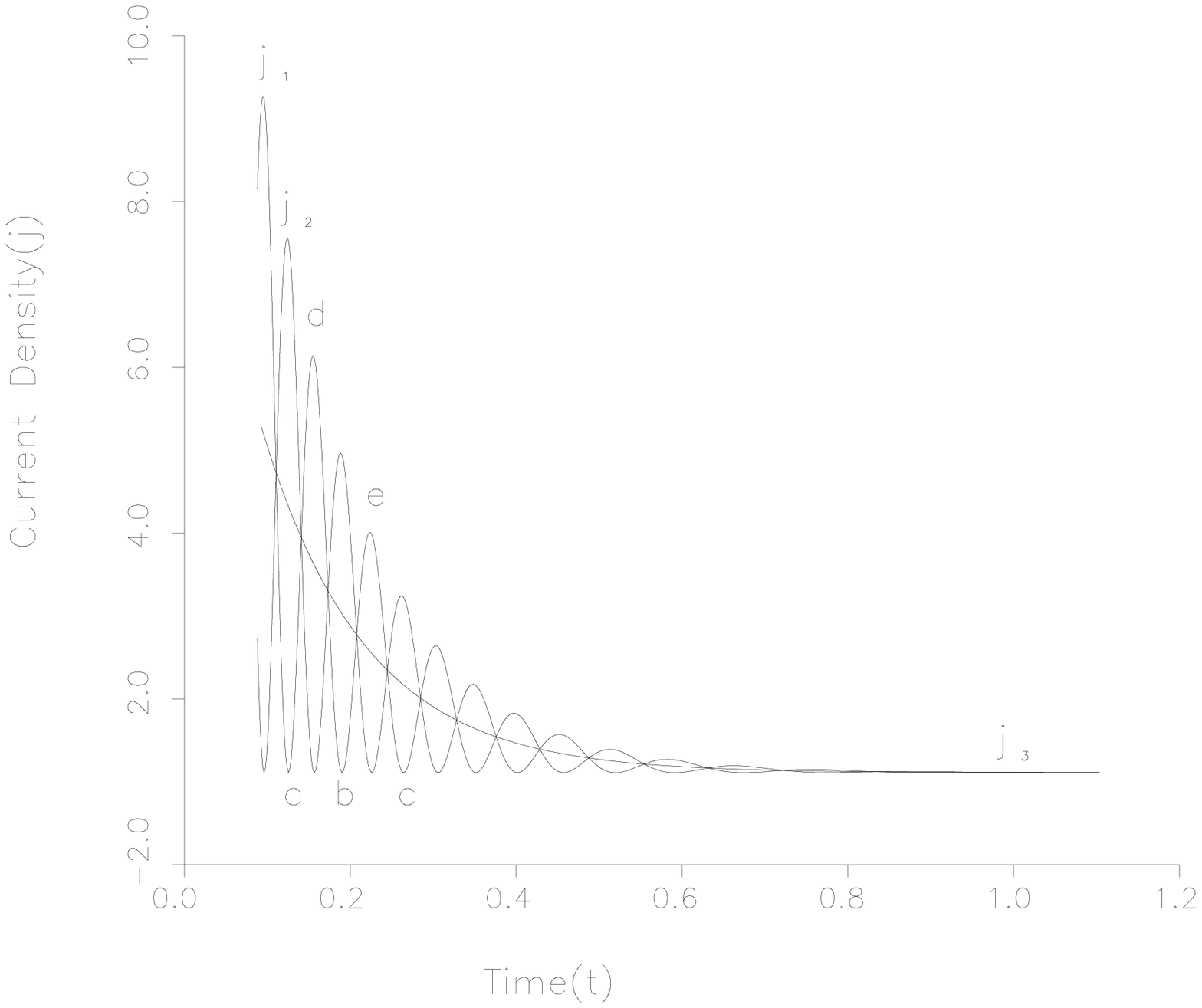,width=7cm,height=6cm}
\vspace{0.3cm}
\caption{A plot of the Current Density(j) versus Time(t) in arbitrary
unit;
(a) for $j_{_{0}}$,in case of both sets of solutions, (b) for first set of 
solutions $\psi_{1}$, $\psi_{2}$, (c) for second set of solutions $\psi_{3}$,
$\psi_{4}$, (d) superposition of figures (b) and (c) in Schr{\"o}dinger
Model and (e) in case of de Sitter spacetime.}
\end{figure}
\end{document}